# SARS-CoV-2 spread and quarantine fatigue: a theoretical model

Ariel Félix Gualtieri*, Pedro Hecht

*Universidad de Buenos Aires. Facultad de Odontología. Cátedra de Biofísica y Bioestadística. Buenos Aires, Argentina*

**Abstract**

During the COVID-19 pandemic, community containment measures have been implemented in different parts of the world. However, despite government recommendations and mandates, a progressive increase in mobility was observed in several sites. This phenomenon is often referred to as "quarantine fatigue". Mathematical models have been used for more than a century to study the dynamics of diseases spread. Since the start of the COVID-19 pandemic, a large number of models have been developed to investigate different aspects of the pandemic. The aim of the present work has been to design and explore a mathematical model of SARS-CoV-2 spread with quarantine fatigue. A deterministic model represented by a system of differential equations, based on Susceptible-Infected-Recovered (SIR) dynamics, was developed. The model was explored by means of computational simulations. The results obtained show that, in some scenarios, after a first prevalence peak, fatigue can cause a second peak. Thus, the dynamic behavior of the model suggests that fatigue may have contributed to the production of the second peak of infected people observed in different parts of the world between 2020 and early 2021. The explorations performed also showed that the model is sensitive to infection rate and other parameters related to propagation dynamics. The developed model is relatively simple and has important limitations for performing accurate predictions in the real world. However, our study highlights the importance of considering quarantine fatigue in the management of the current COVID-19 pandemic, as well as of possible future pandemics caused by other pathogens.

_________

*Author for correspondence. Universidad de Buenos Aires. Facultad de Odontología. Cátedra de Biofísica y Bioestadística. Marcelo T. de Alvear 2142. Piso 17 B. (C1122AAH) Ciudad de Buenos Aires, Argentina. Email: ariel.gualtieri@odontologia.uba.ar

# 1. Introduction

The first case of coronavirus disease-19 (COVID-19), the disease caused by the severe acute respiratory syndrome coronavirus 2 (SARS-CoV-2), was reported in Wuhan city, China, in December 2019 (World Health Organization, 2020). Since then, the pandemic has spread around the globe, and by the end of March 2021 has already produced about 126 million cases and more than two and a half million deaths (Worldometers.info, 2021).

During the COVID-19 pandemic, different health control strategies have been implemented, such as isolation, quarantine, social distancing and community containment (Mackenzie *et al*., 2020). Community containment involves the entire population, and includes measures that limit interaction between people and restrict movement, such as canceling social gatherings, closing schools, closing work offices and restricting travel (Wilder-Smith & Freedman, 2020).

Quarantine and isolation are old measures used by mankind to control the spread of diseases (Gensini *et al*., 2004; Cetron & Landwirth, 2005; Nugent, 2020). However, community containment measures are less common. Previously, community containment measures had already been implemented in some places and, along with quarantine and isolation, were successful in stopping the 2003 SARS outbreak (Wilder-Smith *et al*., 2020). However, the COVID-19 pandemic led to the implementation of community containment measures with unprecedented global reach.

During the COVID-19 pandemic, it has been observed that, despite community containment measures, "quarantine fatigue" occurs in some scenarios, which would cause a consequent increase in cases (Dincer & Gillanders, 2020; Sun *et al*., 2020; Zhao *et al*., 2020). We will say that quarantine fatigue exists when, despite government mandates and recommendations, population mobility increases over time. The study of this phenomenon could be related to the field of behavioral epidemiology, a field where mathematical models would be very useful tools (Bauch *et al*., 2013).

Mathematical models have been used for more than a century to study the dynamics of diseases spread (Brauer, 2017). In compartmental epidemiological models, the population is divided into compartments, according to the different stages of the disease or other characteristics related to the epidemiology of the phenomenon under study. For example, in a classical SIR model, the population is divided into three compartments: susceptible (S), infected (I) and recovered with immunity (R). The transition between different stages is then represented, through a formal scheme, as a function of time. In the classical SIR model, there are two transitions: S→I and I→R. Each transition depends on a parameter. Thus, the transition S→I depends on the infection rate and the transition I→R depends on the recovery rate (Weiss, 2013).

Another important measure for the development of epidemiological models is the basic reproductive number, $R_0$, which is defined as the average number of new infections that are caused by an infected individual within a population where all other individuals are susceptible (Mishra *et al*., 2011). The $R_0$ of SARS-CoV-2 would be quite variable. In a meta-analysis, Alimohamadi *et al*. (2020) found a range of $R_0$ from 1.9 to 6.49, and estimated a pooled $R_0$ of 3.32, with a 95% confidence interval of 2.81 to 3.82.

Since the spread of SARS-CoV-2 began, a large number of models have been developed to study different aspects of the pandemic (Adiga *et al*., 2020; Mustafa *et al*., 2020). Different models show how community containment measures can reduce virus circulation (for example, Zhou *et al*., 2020; Fang *et al*., 2020). The objective of the present work has been to develop and explore a theoretical model of SARS-CoV-2 spread with quarantine fatigue.

## 2. Materials and methods

### 2.1. The model

A deterministic model was developed to simulate the spread of SARS-CoV-2. The model is represented by the following system of 6 differential equations:

$$\frac{dS_H}{dt} = -\frac{\beta_H S_H I_H}{N_H} - fS_H \tag{1}$$

$$\frac{dI_H}{dt} = \frac{\beta_H S_H I_H}{N_H} - \gamma I_H - fI_H \tag{2}$$

$$\frac{dR_H}{dt} = \gamma I_H - fR_H \tag{3}$$

$$\frac{dS_M}{dt} = -\frac{\beta_M S_M I_M}{N_M} + fS_H \tag{4}$$

$$\frac{dI_M}{dt} = \frac{\beta_M S_M I_M}{N_M} - \gamma I_M + fI_H \tag{5}$$

$$\frac{dR_M}{dt} = \gamma I_M + fR_H \tag{6}$$

The population was divided into two groups, which we gave arbitrary names in order to simplify the nomenclature: *at home* (*H*) and *mobile* (*M*). Within group *H*, individuals spend more time at home and are less mobile. In contrast, within group *M*, individuals spend less time at home and are more mobile.

A SIR scheme was considered. Thus, $S_H(t)$, $I_H(t)$ and $R_H(t)$ represent the number of susceptible, infected and recovered individuals within group $H$ at time $t$, respectively. Similarly, $S_M(t)$, $I_M(t)$ and $R_M(t)$ are the number of susceptible, infected and recovered individuals within group $M$. $N_H(t)$ and $N_M(t)$ are the total number of individuals in groups $H$ and $M$, respectively:

$$N_H(t) = S_H(t) + I_H(t) + R_H(t) \tag{7}$$

$$N_M(t) = S_M(t) + I_M(t) + R_M(t) \tag{8}$$

$S_P(t)$, $I_P(t)$ and $R_P(t)$ are the number of susceptible, infected and recovered individuals in the whole population:

$$S_P(t) = S_H(t) + S_M(t) \tag{9}$$

$$I_P(t) = I_H(t) + I_M(t) \tag{10}$$

$$R_P(t) = R_H(t) + R_M(t) \tag{11}$$

$N_P(t)$ is the total number of people in the population:

$$N_P(t) = N_H(t) + N_M(t) \tag{12}$$

Within the population, $i_H(t)$ and $i_M(t)$ are the proportion of infected individuals belonging to groups $H$ and $M$, respectively:

$$i_H(t) = \frac{I_H(t)}{N_P(t)} \tag{13}$$

$$i_M(t) = \frac{I_M(t)}{N_P(t)} \tag{14}$$

The proportion of infected individuals in the whole population is $i_P(t)$:

$$i_P(t) = \frac{I_P(t)}{N_P(t)} \tag{15}$$

In both groups, the *per capita* infection rate or force of infection, $\lambda$, is assumed to depend on the proportion of infected individuals: $\lambda = \beta I/N$, where $\beta$ is an infection parameter proportional to the contact rate and the probability of transmission (Keeling & Rohani, 2008).

Thus, $\beta_H$ and $\beta_M$ are the infection parameters in groups H and M, respectively. We assume that $\beta_H<\beta_M$. Infected individuals recover and acquire immunity with a recovery rate $\gamma$.

After a certain tolerance period T, individuals in group H begin to fatigue. In terms of the model, this means that they move from group H to group M with a fatigue rate $f$. Thus, from time instant T onwards, the number of individuals in group H decreases, while the number of individuals in group M increases. The diagram of the model is shown in Figure 1.

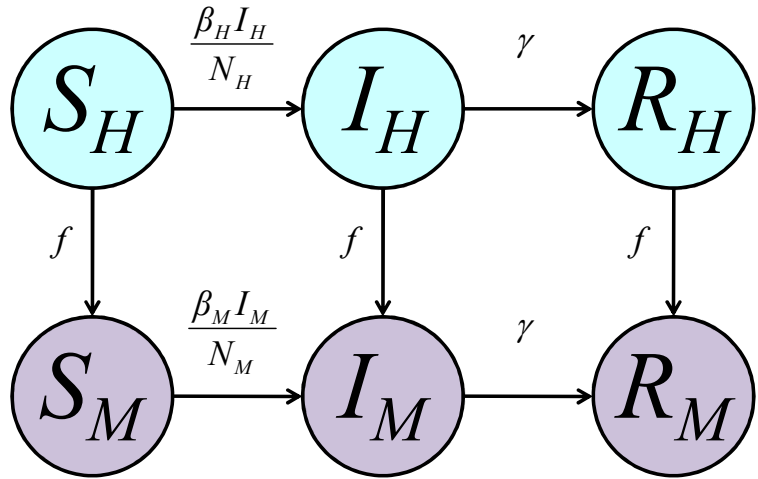

**Figure 1. Schematic flow chart of the model.**

## 2.2. Numerical simulations

The dynamics of the model were explored through numerical simulations. $\beta$ values of between 0.06 and 0.30 1/day were used, in agreement with the range of reproductive numbers published in previous meta-analyses (Alimohamadi *et al.*, 2020; Billah *et al.*, 2020). We assumed $R_0=\beta/\gamma$ (Keeling & Rohani, 2008). An infectious period of 14 days was established (Martelucci *et al.*, 2020). The recovery rate, $\gamma$, is the inverse of the infectious period (Earn, 2008), so a value of 0.07 1/day was used for $\gamma$. The fatigue rate $f$, the ratio $\beta_H/\beta_M$, the tolerance period T and the initial conditions were arbitrary.

A benchmark case was established with the following initial conditions and parameters. Initial conditions (number of individuals): $S_H(0)=990000$, $I_H(0)=10000$, $R_H(0)=0$, $S_M(0)=99000$, $I_M(0)=1000$, $R_M(0)=0$. Parameters (1/day): $\beta_H=0.08$, $\beta_M=0.20$, $\gamma=0.07$, $f=0.06$. A T period of 200 days was considered.

The effects of $\beta_H$, $\beta_M$, $f$ and $T$ on the temporal evolution of the prevalence in the whole population, $i_P(t)$, were studied. These measures were modified one at a time, keeping the rest of the conditions as in the benchmark case. The results were analyzed qualitatively.

Simulations and plots were performed using R software, version 4.0.3 (R Core Team, 2020). The R *deSolve* package was used for the simulations (Soetaert *et al.*, 2010). The plots were performed with R base packages and R *ggplot2* package (Wickham, 2016).

## 3. Results

Figure 2 shows the evolution of the proportion of infected individuals in the benchmark case. The whole population prevalence curve, $i_P(t)$, has 2 noticeable peaks. The first peak occurs around 40 days, with a prevalence of approximately 0.04. After the first peak, the prevalence drops and remains at relatively low values until the 200-day tolerance period $T$ ends. From day 200, $i_P(t)$ begins to increase again and a second peak occurs around 270 days, with a prevalence of approximately 0.12. After this second peak, the prevalence in the population decreases again and tends to zero.

The fraction $i_M(t)$ also experiences two peaks, which coincide with the two peaks of whole population prevalence. This indicates that the two peaks of whole population prevalence are produced by the increase of infected in group $M$. After the first peak, $i_M(t)$ drops to values very close to zero, and remains almost zero until it begins to increase after day 200.

The $i_H(t)$ fraction grows more slowly than the $i_M(t)$ fraction, and has a relatively low peak of about 0.015 at around 110 days. The $i_H(t)$ fraction prevents $i_P(t)$ from reaching near zero between the two peaks of whole population prevalence.

In Figure 3 it is shown the evolution of the number of susceptible, infected and recovered individuals within each group, $H$ and $M$, which clarifies the results presented in Figure 2. Within group $H$, after the 200-day tolerance period $T$, there are about 12 000 infected individuals, but there are still about 780 000 susceptible people. Infection of these susceptible individuals causes the second peak of whole population prevalence.

Fatigue contributes to the second peak being higher than the first, since, due to the fatigue rate, after 200 days, $H$ individuals gradually become $M$ individuals, and susceptible individuals in the group $M$ are more likely to become infected than susceptible individuals in the group $H$.

The influence of $\beta_H$, $\beta_M$, $f$ and $T$ on the total proportion of infected people, $i_P(t)$, is depicted in figures 4 to 7, respectively. As $\beta_H$ increases, the first peak of population prevalence increases, and the prevalence values between the two peaks are also higher (Figure 4). However, when $\beta_H$ increases, the second peak prevalence is reduced. This is because increasing $\beta_H$ reduces the number of susceptible individuals remaining in group $H$ after 200 days.

When $\beta_M$ increases, the two prevalence peaks increase and occur earlier (Figure 5). The increase in fatigue rate $f$ does not influence the first prevalence peak, but, as expected, causes an increase in the second peak (Figure 6). The increase in $T$ reduces the second prevalence peak and delays its arrival (Figure 7). In particular, when there is no tolerance period ($T=0$), the curve presents a single peak, around day 50.

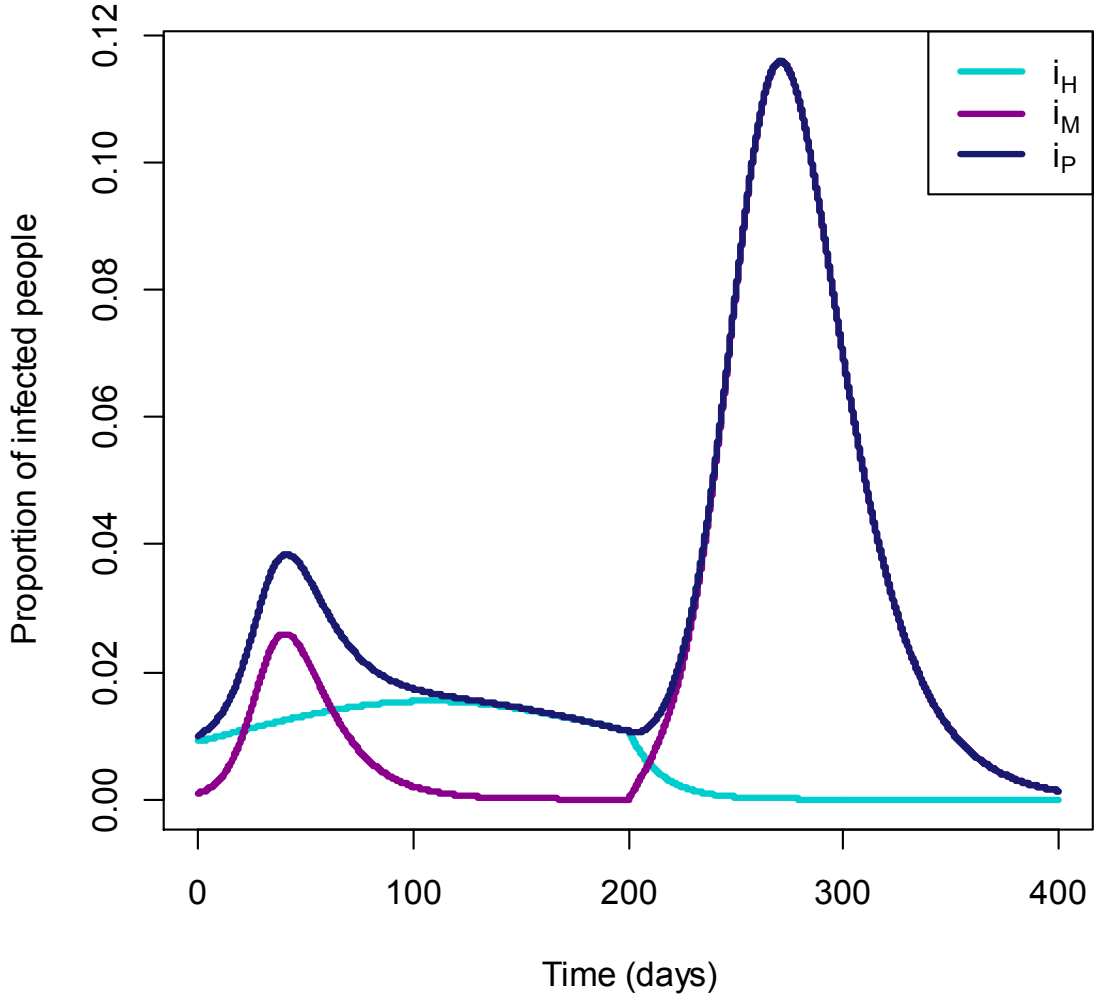

**Figure 2. Evolution of prevalence in the benchmark case. Proportions of infected individuals belonging to groups $H$ ($i_H$) and $M$ ($i_M$), and prevalence in the whole population ($i_P$). Initial conditions: $S_H(0)=990\,000$, $I_H(0)=10\,000$, $R_H(0)=0$, $S_M(0)=99\,000$, $I_M(0)=1\,000$, $R_M(0)=0$. Parameters (1/day): $\beta_H=0.08$, $\beta_M=0.20$, $\gamma=0.07$, $f=0.06$. $T=200$ days.**

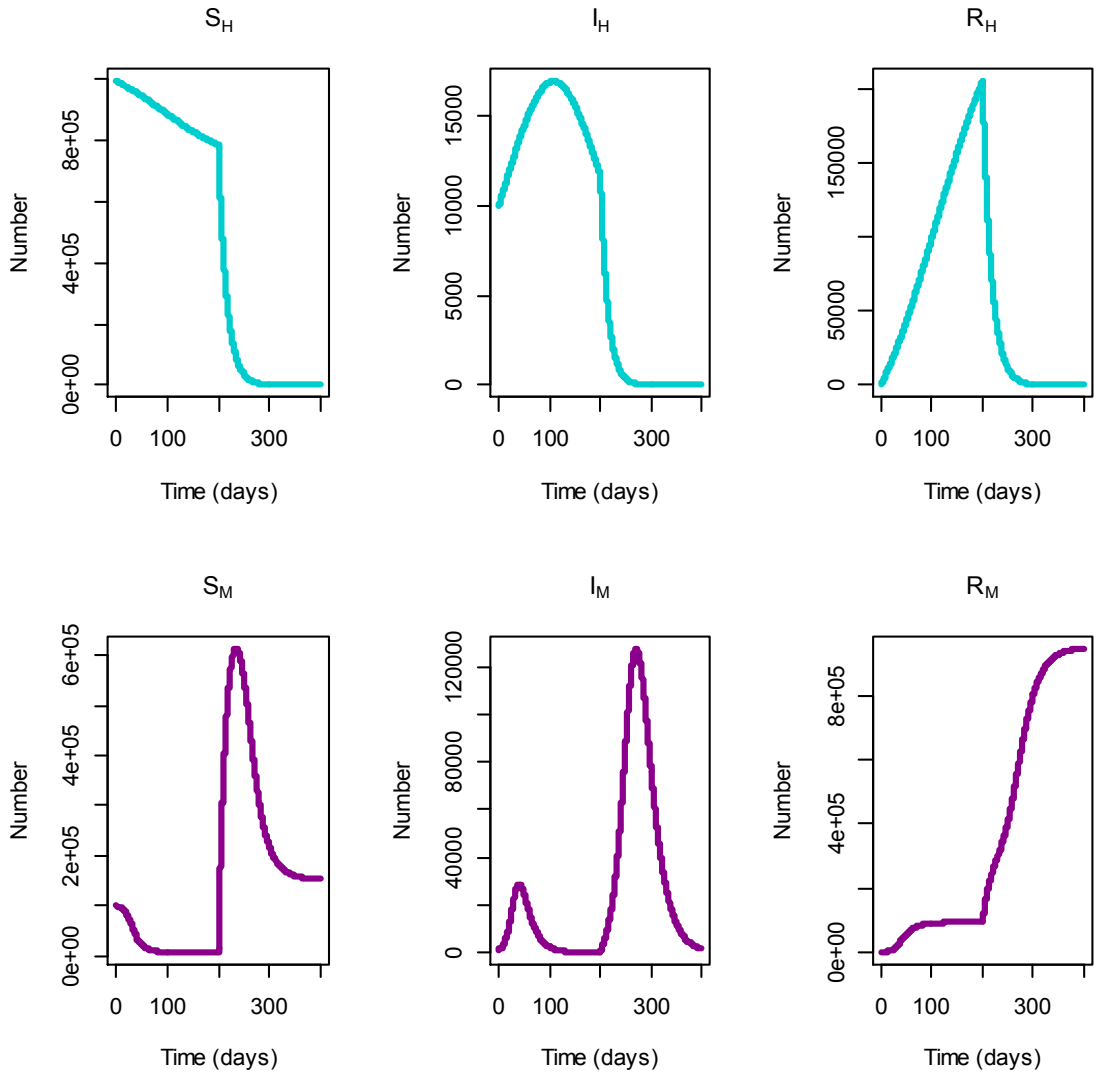

**Figure 3.** Evolution of susceptible (*S*), infected (*I*) and recovered (*R*) individuals in the benchmark case, groups *H* and *M*. Initial conditions: $S_H(0)=990\,000$, $I_H(0)=10\,000$, $R_H(0)=0$, $S_M(0)=99\,000$, $I_M(0)=1\,000$, $R_M(0)=0$. Parameters (1/day): $\beta_H=0.08$, $\beta_M=0.20$, $\gamma=0.07, f=0.06$. $T=200$ days.

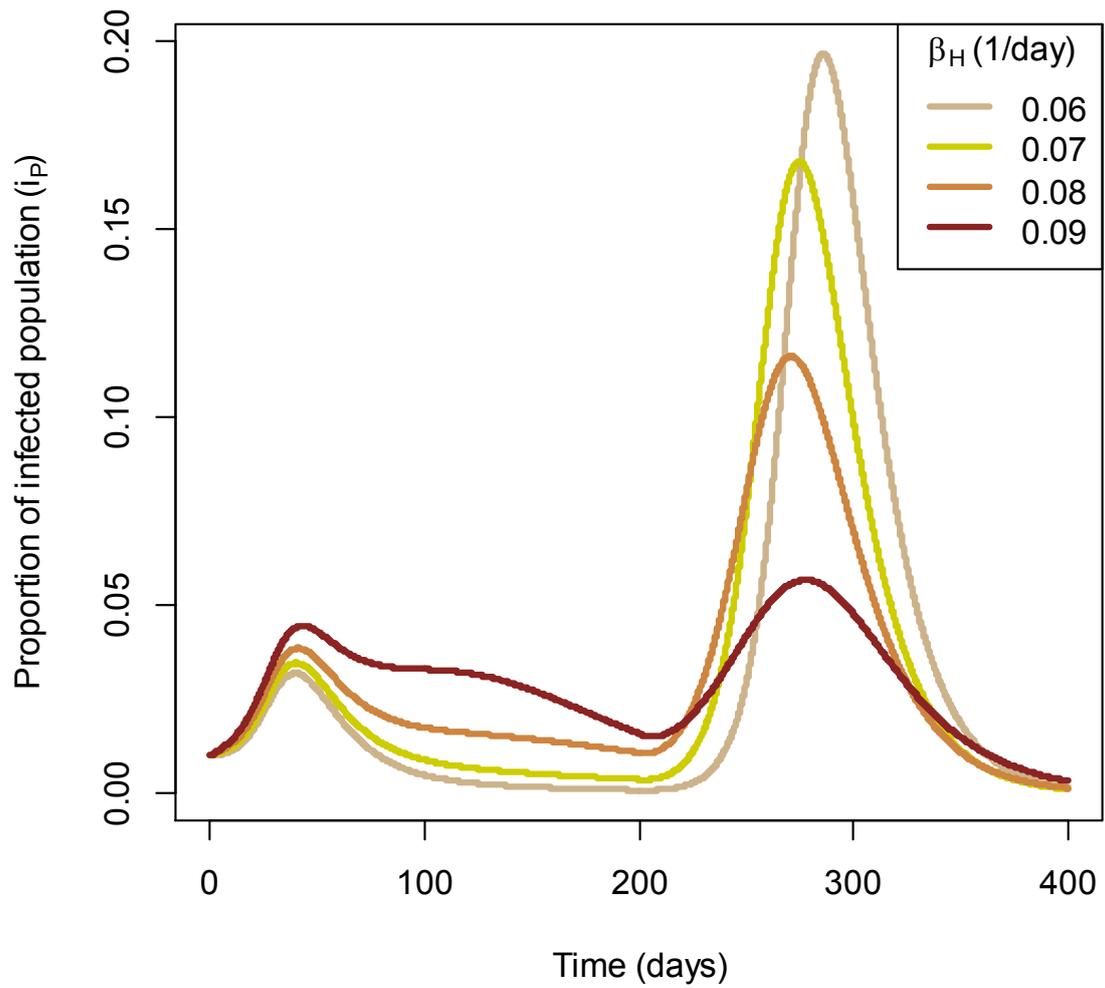

**Figure 4. Influence of the transmission parameter within group $H$ ($\beta_H$) on the evolution of the prevalence in the whole population ($i_P$). $\beta_H$ (1/day)={0.06, 0.07, 0.08, 0.09}. The rest of the conditions were maintained as in the benchmark case.**

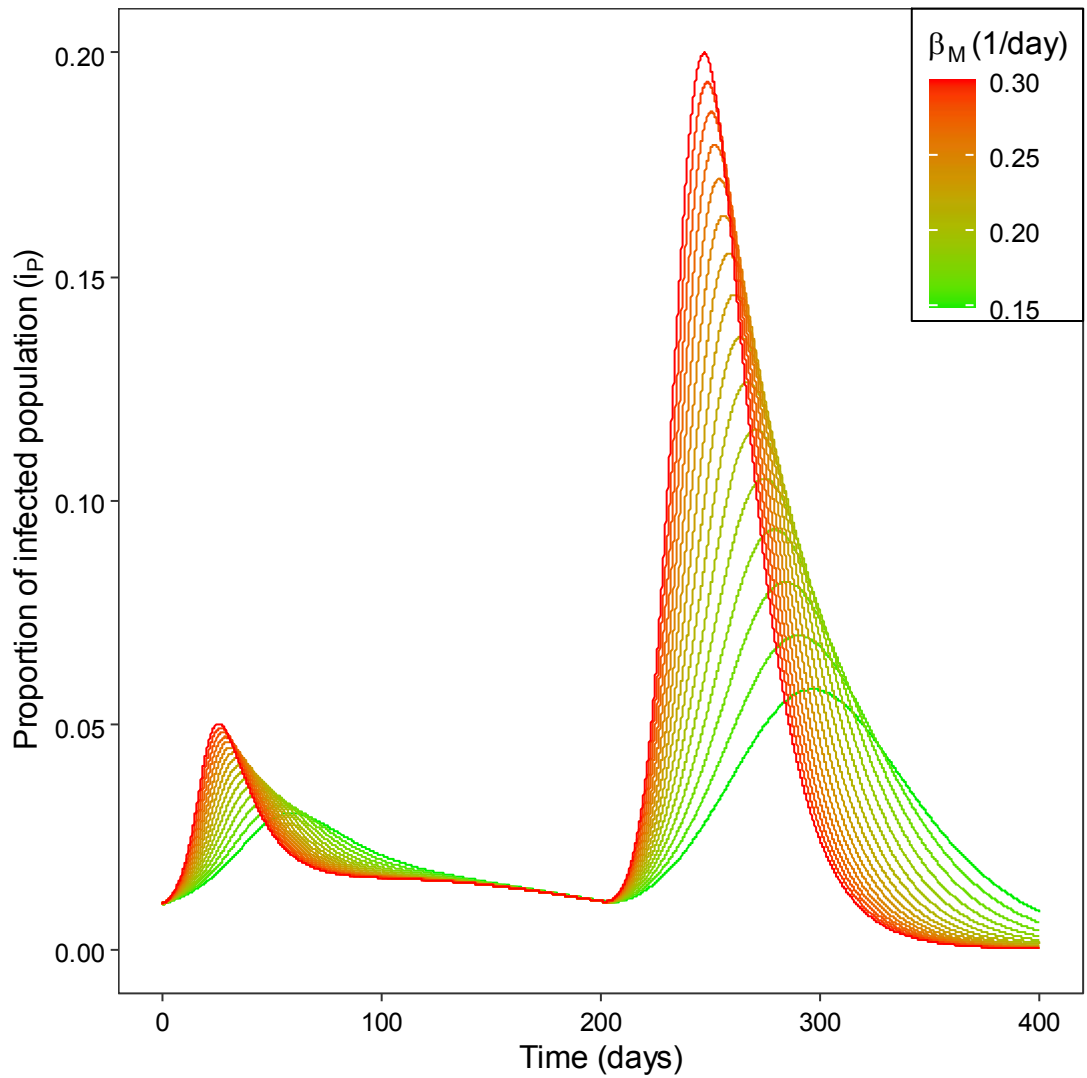

**Figure 5. Influence of the transmission parameter within group *M* ($\beta_M$) on the evolution of the prevalence in the whole population ($i_P$). $\beta_M$ (1/day)={0.15 to 0.30, by increments of 0.01}. The rest of the conditions were maintained as in the benchmark case.**

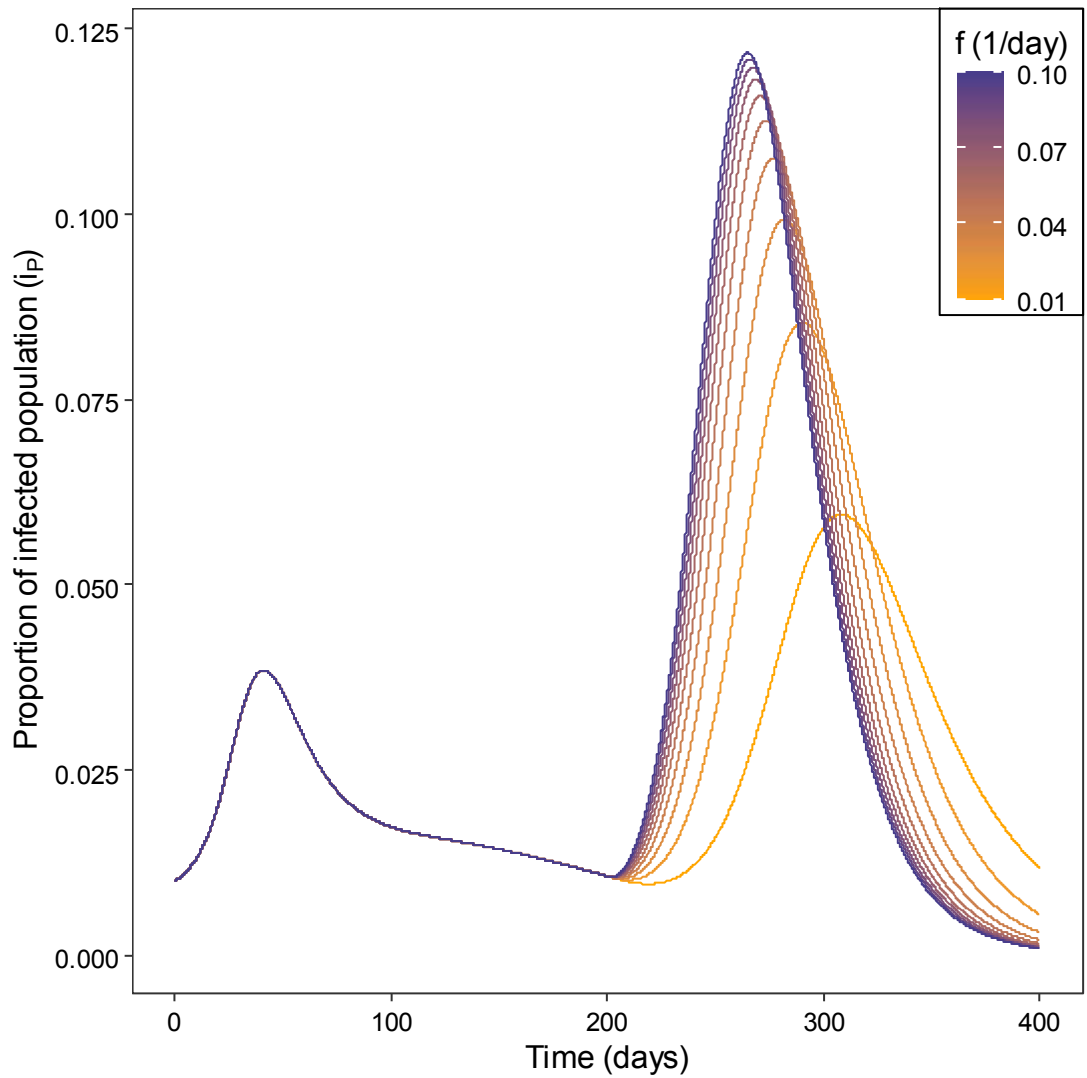

Figure 6. Influence of the fatigue rate *f* on the evolution of the prevalence in the whole population ($i_P$). *f* (1/day)={0.01 to 0.10, by increments of 0.01}. The rest of the conditions were maintained as in the benchmark case.

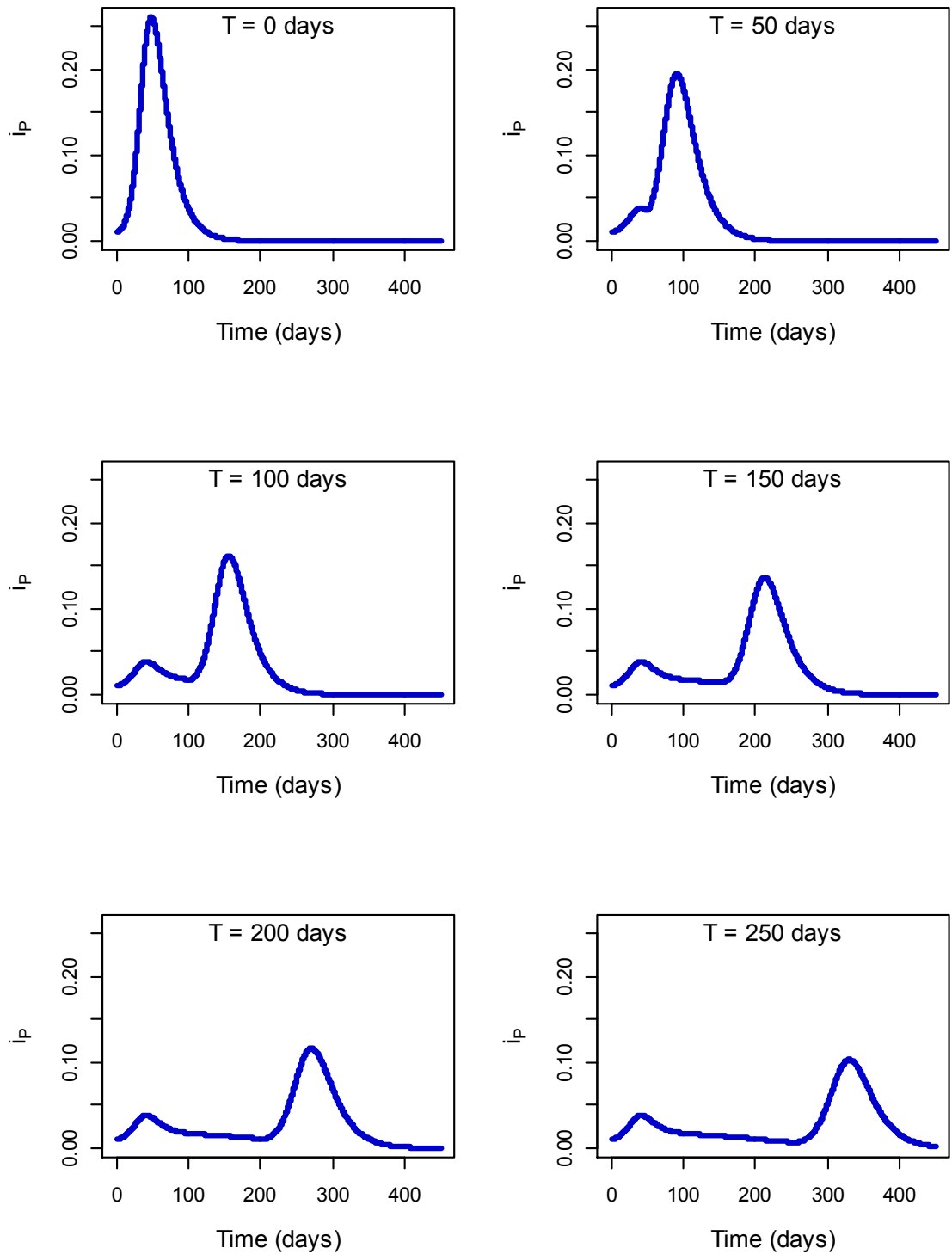

**Figure 7. Influence of the tolerance period *T* on the evolution of the prevalence in the whole population (*i*<sub>P</sub>). *T* (days)={0 to 250, by increments of 50}. The rest of the conditions were maintained as in the benchmark case.**

# 4. Discussion

In a pandemic caused by a novel and aggressive virus, such as SARS-CoV-2, in the absence of other tools, community containment measures are very useful, as they reduce infections and deaths, and prevent the collapse of health systems (Prem *et al*., 2020; Salzberger *et al*., 2020). However, they could also produce negative effects in other areas (Hawryluck *et al*., 2004; Chu *et al*., 2020; Dubey *et al*., 2020). The challenge is to find an optimal solution that takes into account the effects of different variables (Chowdhury *et al*., 2020). Mathematical models could be powerful resources to address such challenge.

In the present work, we have developed a model that attempts to represent quarantine fatigue in the context of the pandemic produced by the spread of the SARS-CoV-2 virus. Through numerical simulations, we show that the model is sensitive to parameters related to the infection rate, the tolerance time to community containment measures and the fatigue rate.

Between 2020 and early 2021, in different parts of the world, the evolution of the number of infected people over time has shown two notable peaks (World Health Organization, 2021). The simulations performed in our study show how, within the model we have proposed, fatigue can cause a second prevalence peak.

The relaxation of protective measures and its negative impact on epidemics control would not be something new. This phenomenon could also occur with some sexually transmitted diseases (Martí-Pastor *et al*., 2015), and could have occurred in some scenarios during the influenza A (H1N1) pandemic in 2009 (Herrera-Valdez, 2011).

One of the uses of an epidemiological model is to alert about the information that should be collected in order to have more knowledge about the spread and control of the disease under study (Anderson & May, 1998). In this sense, we consider that our model could stimulate the development of field work focused on solving important questions related to quarantine fatigue, such as the following: when does fatigue begin?, what situations promote or discourage it?, how is it distributed within the population?

Epidemiological models enable the study of complex phenomena by simplifying reality. Depending on each model, this approach may have various limitations. Our model, which is relatively straightforward, presents different simplifications that would limit its usefulness. Some of the most important ones are mentioned below.

The model assumes that there are only two groups in relation to the degree of mobility. Individuals from one group do not mix with individuals from the other group. The flow between the two groups is unidirectional: individuals with less mobility, when fatigued, can move to the group with higher mobility, where the infection rate is higher. But individuals with higher mobility can never move to the other group. Infection rates within each group do not change as a function of time. No distinction is made between symptomatic and asymptomatic infected individuals. There is no division by age group.

Protective behaviors, such as the use of masks, are not explicitly represented. The model does not include testing, isolation nor vaccination strategies. Different strains of the virus are also not considered.

However, our model does not pretend to represent a specific scenario, nor to make accurate predictions. Instead, it is a theoretical model that would serve as a first approach to the study of quarantine fatigue. This model could be taken as a scaffold to build a more complex and realistic model, focused on particular cases.

## 5. Conclusion

The present work suggests that fatigue may have contributed to the production of the second peak of infected people observed in different parts of the world between 2020 and early 2021. But our model is very simple, which makes it have important limitations to be applied in real scenarios. Although the model was developed for the COVID-19 pandemic, it could be the starting point for studying quarantine fatigue in potential future pandemics.

## Declaration of competing interest

The authors declare that they have no conflict of interest.

## Acknowledgments

The authors gratefully acknowledge Universidad de Buenos Aires for providing financial support (grant numbers UBACyT 20020160100118BA, UBACyT 20020190200179BA).